# Van der Waals Materials for Atomically-Thin Photovoltaics: Promise and Outlook

*Deep Jariwala[1,2], Artur R. Davoyan[1,2,3], Joeson Wong[1] and Harry A. Atwater[1,2,3,4]\**

[1]Department of Applied Physics and Material Science, California Institute of Technology, Pasadena, CA-91125, USA

[2]Resnick Sustainability Institute, California Institute of Technology, Pasadena, CA-91125, USA

[3]Kavli Nanoscience Institute, California Institute of Technology, Pasadena, CA-91125, USA

[4]Joint Center for Artificial Photosynthesis, California Institute of Technology, Pasadena, CA-91125, USA

**Abstract:** Two-dimensional (2D) semiconductors provide a unique opportunity for optoelectronics due to their layered atomic structure, electronic and optical properties. To date, a majority of the application-oriented research in this field has been focused on field-effect electronics as well as photodetectors and light emitting diodes. Here we present a perspective on the use of 2D semiconductors for photovoltaic applications. We discuss photonic device designs that enable light trapping in nanometer-thickness absorber layers, and we also outline schemes for efficient carrier transport and collection. We further provide theoretical estimates of efficiency indicating that 2D semiconductors can indeed be competitive with and complementary to conventional photovoltaics, based on favorable energy bandgap, absorption, external radiative efficiency, along with recent experimental demonstrations. Photonic and electronic design of 2D semiconductor photovoltaics represents a new direction for realizing ultrathin, efficient solar cells with applications ranging from conventional power generation to portable and ultralight solar power.

*Corresponding author: haa@caltech.edu

Since the isolation of graphene as the first free-standing two-dimensional (2D) material (from graphite), the class of layered 2D materials with weak van der Waals inter-planar bonding has expanded significantly. Two-dimensional materials now span a great diversity of atomic structure and physical properties. Prominent among these are the semiconductor chalcogenides of transition and basic metals (Mo, W, Ga, In, Sn, Re etc.)[1-3], as well as layered allotropes of other p-block elements of the periodic table such as P, As, Te etc.[4] The availability of atomic layer thickness samples of stable, passivated, and dangling bond free semiconductor materials ushers in a new phase in solid state device design and optoelectronics.[1, 5-8] A notable feature of the metal chalcogenide 2D semiconductors is the transition from an indirect bandgap in bulk to direct bandgap ($E_g$) in monolayer form, resulting in a high photoluminescence quantum yield (PL QY)[9-10] in turn corresponding to high radiative efficiency. This combined with the bandgap ranging from visible to near infrared part of the spectrum (1.1 to 2.0 eV)[1] makes the chalcogenides attractive as candidates for photovoltaic applications. Likewise, small bandgap elemental 2D semiconductors such as black phosphorus and alloys of arsenic and phosphorus are attractive candidates for thermo-photovoltaic applications in their few layer to bulk form ($E_g \leq 0.6$ eV). In the ultrathin limit (<5 layers) the increase in bandgap to > 1 eV due to quantum confinement makes them attractive for conventional photovoltaic applications as well.[11-13] However, the current inability to synthesize these materials in a scalable manner with precise control over thickness combined with their lack of air stability has restricted their investigation to preliminary photocurrent generating devices from mechanically exfoliated samples.

Owing to the above-mentioned attributes, 2D semiconductors have been used to demonstrate devices with photovoltaic effects.[14-16] In most cases these have been proof-of-concept devices reporting the basic feasibility of photovoltage generation in such material systems. At the same time, the field of photovoltaics is at an advanced stage, with GW-scale commercial production for silicon-based photovoltaics now a reality and cost of solar power expected to achieve parity with fossil fuel based power plants by 2020. Therefore, simply the demonstration of a photovoltaic effect in novel and emerging semiconductors is no longer by itself a requisite for sustained interest from the perspective of photovoltaics application. In this perspective, we argue that 2D semiconductors are indeed promising for photovoltaic applications and have the potential to not only match but also surpass the performance and complement the conventional photovoltaic technologies based on Si and GaAs. We present a detailed and comparative analysis based on optical and electronic properties of 2D semiconductors and conclude that it is feasible in principle to design photovoltaic devices with power conversion efficiencies exceeding 25%. We further present strategies for photonic and electronic device design to maximize light trapping/useful absorption and efficient extraction of photo-excited carriers, respectively. We also give a brief outlook for the prospects for tandem integration to enable a '2D-on-silicon' dual junction tandem solar cell. Finally, we provide a roadmap for the advances needed to achieve high-performance photovoltaic devices with nanometer thick absorbers and provide a critical assessment for future research developments in this area.

**Absorption and Photonic Design:**

Light absorption in the active layers of a photovoltaic cell is one of the key performance metrics that dictates device efficiency. For semiconductors, including 2D materials, the absorption is governed by the electronic band structure and energy bandgap. There is an inherent trade-off between bandgap (and voltage) with absorption (and photocurrent). In Fig. 1a, we summarize the bandgap energies and absorption coefficients for all major photovoltaic materials investigated to date at the commercial and research scales. As is evident from Fig. 1a and the discussion above, most materials considered for photovoltaic applications have energy bandgaps close to the optimum value of 1.34 eV.

In modern photovoltaic devices, light trapping techniques are often employed to maximize incident light absorption and photocurrent generation in the active layer to increase the cell efficiency, which also has the benefit of reducing thickness and thus material use and device weight.[17-19] The extent of light trapping in a medium in both the ray optic (bulk)[20] and nanophotonic (sub-wavelength) regimes[21] may be quantified by the ratio of imaginary and real parts of dielectric function, i.e., loss tangent. Figures 1b and 1c show spectral dependence of absorption coefficients and loss tangents in the sulfides and selenides of Mo and W compared with Si, GaAs, and the recently emerging hybrid organic-inorganic perovskites[22-23]. Owing to the high refractive indices of the transition metal dichalcogenides (TMDCs), they exhibit significantly higher absorption per unit thickness as compared to Si, GaAs, and even the perovskites. Thus, TMDCs are ideally suited for highly absorbing ultrathin photovoltaic devices.

Despite the large absorption coefficient values for TMDCs, a free-standing monolayer only absorbs ~10 % of the above bandgap, normally incident photons[9, 24-25] (Figure 2a). In multilayers with bulk-like electronic structure up to ~25 nm thick, the absorption is less than 40%,[26] and high surface reflectivity limits absorption. Therefore, light trapping designs and device architectures will play a critical role in enabling efficient 2D semiconductor photovoltaics in the ultrathin limit. Several strategies have been proposed and preliminary demonstrations have been reported including use of plasmonic metal particles, shells or resonators to enhance photocurrent and photoluminescence.[27-37] More sophisticated and lossless dielectric optical cavities such as photonic crystals and ring resonators have also been used to enhance absorption, mainly aimed at emission applications in monolayer samples.[38-41]

For large area photovoltaic applications, light trapping strategies that involve little or no micro- or nanofabrication and patterning are desirable to improve performance while minimizing cost.[42-43] An interesting strategy towards this end is the use of thin film interference. Figures 2b and 2d show two common strategies for lithography-free absorption enhancement in ultrathin semiconductor films. In both of these designs, a highly reflective metal (e.g., Au or Ag) is used as a part of an "open cavity", such that resonant light trapping conditions are met. Hence, even an atomically thin absorber spaced $\lambda/4$ away from a reflector (Figure 2b) enables destructive interference at the interface, resulting in significant absorption enhancement.[44-45] Another approach is to place an ultrathin absorbing layer in a direct contact with the back reflector[46-47]. This strategy has worked well for TMDC devices with Au and Ag back reflectors, resulting in record broadband absorption (> 90%)[26] and quantum efficiency (>70%) values[48] in < 15 nm thick active layer devices. Note that back reflectors are widely employed in conventional thin-film photovoltaic devices, where absorption enhancement is due to multi-pass light interactions within the semiconductor. We also note that the light trapping may be further enhanced by the use of nanostructured resonators coupled to thin film absorbers,[49-50] as is shown schematically in Figures 2c and 2e.

Overall, due to the large values n and $\alpha$ for TMDCs, trapping nearly 100% of the incident light may be achieved for few-nm thick active layers. However, enabling broadband, nearly perfect absorption in sub-nm thick monolayers is more challenging and less scalable, as currently-identified light trapping approaches are likely require fabrication of nanoscopic resonators or antennas on top of or etched into the monolayer.[51-53]  Thus an open challenge at present for atomically thin photovoltaics is the photonic design of nanostructures that retain the electronic structure and photonic properties of monolayer 2D materials, while also exhibiting optical cross sections equivalent to multilayer/bulk samples.

**Carrier Collection and Electronic design:**

While photon absorption represents one limit to photovoltaic efficiency, the collection of photo-excited charge carriers dictates the practical limitations on the maximum attainable current. Carrier collection is often quantified in terms of the probability of carrier collection per absorbed photon, often termed the internal photocarrier collection efficiency (IPCE) or internal quantum efficiency (IQE), and is sensitive to several factors, including device dimension and design, contacts and material quality. Frequently also reported is the external quantum efficiency (EQE) or probability of carrier collection per incident photon, which can asymptotically approach the IQE for perfect absorption. In reality, the device design and material quality plays a key role in determining the IQE, whereas EQE depends on these factors as well as photonic design.

A 2D semiconductor absorber can be integrated into a photovoltaic device in several possible ways. Figure 3 details the schematic designs of some possible approaches. To separate the photo-excited carriers, one either requires a built-in potential within the absorber layer, often provided by a pn junction, or alternatively a uniformly doped absorber layer cladded by carrier selective contacts.[54] This pn junction can either have an in-plane configuration spanning adjacent regions of different composition or doping in the covalently bonded 2D plane, or an out-of-plane configuration featuring vertically stacked van der Waals-bonded layers perpendicular to the 2D plane. In each case, the device structure can consist of a homojunction[55-57] or a heterojunction[58-61] design. Each of these concepts has advantages and limitations. As an example, for vertically stacked pn junctions, 2D semiconductors are uniquely positioned to achieve high QE[37, 48, 62] owing to their atomic-scale thicknesses, ensuring < 10 nm excited carrier transit distances (Figure 3 b). Similarly, in-plane collection devices are well suited for forming junctions via substitutional,[58-59] chemical,[55] thickness variation[63-65] or electrostatic doping[66-69] , as in Figure 3 a, that can enable large open circuit voltages. Vertically-stacked junctions may also be more suitable for multilayer thick absorbers while lateral junctions may be more suitable for monolayer absorbers to maximize shunt resistance and avoid electrical shorts in the device. Likewise, it is also likely to be easier to integrate 2D semiconductor photovoltaics as the component sub-cells of a tandem photovoltaic structure integrated with or on conventional Si[70-71], thin film CIGS, CdTe, or GaAs[72] photovoltaics, or even organic semiconductors[5, 73-74] , where the 2D semiconductor forms a van der Waals vertically stacked device. By contrast, a lateral junction would require in-plane integration of dissimilar materials. For lateral junctions, the absorber layer crystalline quality and minority diffusion length are critical, since carriers must be transported in-plane before reaching the contacts.

The junction design and junction type also dictates the configuration of contacts required for carrier collection. For vertically stacked junctions, one transparent, low-absorptive loss contact is essential for efficient optical absorption. Graphene has emerged as one alternative[75-76]; however, the sheet resistance of graphene still remains comparatively higher than for transparent conductive oxides. Metal contacts are attractive alternatives, particularly for lateral junction devices but metals typically result in loss of active area due to shadowing effects. Nonetheless, with appropriate photonic design one can achieve effectively transparent contacts composed of metallic structures.[77] Carrier selective contacts are also highly desirable for ultrathin 2D absorber layers, where the excited-carrier transit distances are much less than the characteristic carrier diffusion lengths, enabling device design without a built-in potential or electric field to separate carriers within a nearly intrinsic absorber layer. To date, little knowledge or effort has been devoted to the design, optimization or demonstration of carrier selective contacts for 2D TMDC based photovoltaics, and this is an opportunity for further research.

**Progress, Challenges, and Outlook:**

While stable semiconducting TMDCs have only been isolated and studied since 2011, scientific progress has been rapid and extensive. However, a majority of the scientific progress has been achieved using mechanically exfoliated 2D semiconductor layers which have allowed small prototype devices to be realized, but this synthesis method is not scalable to areas of relevance for large scale photovoltaics. Significant effort has also been devoted to large area synthesis of 2D semiconductors via chemical vapor deposition (CVD).[78-79] However, it is only recently that the community has begun to develop an understanding of the issues pertaining growth, defects, and material quality using this method.[80-81] Therefore, even though numerous results have been published demonstrating proof of concept photovoltaic devices, no systematic attempts have been made to address the fundamental issues that underlie development of efficient photovoltaics, i.e. optical absorption, carrier collection, and open-circuit voltage. While several initial concepts for light management have been proposed for atomically thin semiconductors, including the concepts noted above, few approaches have immediate promise for integration into functional devices, and even fewer have the potential cost-effectiveness and scalability. One promising approach is the use of few-layer thickness TMDCs directly placed on reflective metal substrates as highlighted in Figure 2d above. This approach avoids any micro/nanofabrication requirements for enhancing absorption. Further, the 2D TMDC absorber can be directly grown on the metallic substrates over large areas, suggesting a potentially scalable fabrication approach.

To further assess the viability of 2D semiconductor photovoltaics, it is worth evaluating them i) in the context of commercial, mass-produced single junction photovoltaic technologies and ii) to consider 2D semiconductor photovoltaics relative to detailed balance efficiency limits.[82] Figure 4a is a modified detailed balance model comparison of the maximum efficiency for a single junction photovoltaic cells as a function of the absorber layer bandgap, for different values of external radiative efficiency (ERE). In modified detailed balance models, ERE describes the fraction of total recombination current that results in radiative emission that ultimately escapes from a photovoltaic cell, and is assumed to have values ranging from much less than unity up to unity. External radiative efficiency is a function of several parameters, including intrinsic parameters such as material quality and electronic band structure, as well as extrinsic factors such as electronic and photonic design. Similarly, internal radiative efficiency (IRE) represents intrinsic material parameters and describes the fraction of recombination that is radiative *internally* within a photovoltaic device – a closely related concept to the figure-of-merit known as photoluminescence quantum yield used for light emitters. In the asymptotic limit of perfect device design, the maximum ERE achievable is bounded by the IRE. Photovoltaic cells that reach the thermodynamic detailed balance efficiency limit for their bandgaps must have EREs approaching unity, but this is difficult to achieve in practice.[83] Direct bandgap materials such as GaAs can exhibit EREs in the range of 1% < ERE < 20%, as compared to the typically <1% ERE achievable in an indirect bandgap material such as Si. Notably, organic–inorganic hybrid perovskites are direct bandgap materials that have the potential for external radiative efficiencies comparable to those for the highest-quality direct bandgap semiconductors. In the 2D materials literature, ERE is not a commonly reported parameter and instead PLQY is generally reported. By assuming the PLQY to be approximately equal to the IRE, and therefore the maximum achievable ERE as the PLQY, we estimate the efficiency limits of TMDC-based photovoltaic devices as shown in Figure 4a. Monolayer materials with direct bandgaps have recently been shown to exhibit much higher PLQY ( ~ 95% experimentally achieved)[10] and thus also consequently ERE, compared to their indirect bandgap multilayer counterparts. Monolayer 2D semiconductors have relatively larger bandgap values (1.6-2.1 eV) and large exciton binding energies (0.6-0.9 eV)[84-86] due quantum carrier confinement. Large exciton binding energies are *a priori* a disadvantage for high photovoltaic efficiency, and thus 'exciton management' is likely to be an important aspect for 2D semiconductor photovoltaics. For single-absorber devices, maximum attainable power conversion efficiencies in monolayer absorber devices are comparable to those for devices with multilayer absorber layers, which have more optimal bandgap values (1.1-1.3 eV) albeit with low PLQYs($\sim 10^{-4}$–$10^{-2}$) and therefore low ERE due to their indirect bandgap nature. This suggests that although monolayer TMDCs are exciting for photovoltaic power due to their direct bandgaps, even the highest quality monolayer materials with PLQY $\sim$ 1 would only result in an overall detailed balance power conversion efficiency between 26-27% in single-junction devices, which can also be achieved with multilayer TMDCs which have PLQYs values that are 2-3 orders of magnitude lower. The above point is especially relevant, since for multilayer (10-15 nm) TMDCs broadband, angle insensitive light-trapping, efficient carrier collection and device fabrication are relatively straightforward and have been experimentally achieved to a large extent, in contrast to the situation for monolayer absorber layers. However, the bandgaps of monolayer TMDCs are in the range that would be nearly ideal for top cell structures in a two-junction tandem device together with e.g., a Si bottom cell device (Figure 4b). Monolayer photovoltaics might also be interesting for narrow-band light harvesting for colored and semi-transparent photovoltaics in architectural and indoor applications[87] , and also applications where light weight or portability is highly desirable.

To date, power conversion efficiencies in ultrathin 2D semiconductor photovoltaic devices have remained below 5 %, as shown in Figure 4 c. The vast majority of reports of 2D semiconductor photovoltaic device demonstrations have used monolayer absorber layers. However, there are very few quantitative reports of power conversion efficiency under 1 sun AM1.5 or monochromatic illumination, spectral dependence of EQE and absorption in the active layers of the device. This lack of information makes it very challenging to compare literature reports and complicates the assessment of quantitative performance estimates for a reported photovoltaic device. The plots in Figure 4 c show a nearly linear dependence of power conversion efficiency on the external quantum efficiency for bandgap values ranging from 1.1-2 eV and ERE values ranging from 1 to $10^{-4}$. A key observation from this plot is that one can attain greater than 20% power conversion efficiencies, even with bulk-like TMDC absorber layers, provided that the absorption and EQE are nearly perfect, enabled by appropriate photonic and electronic design.

Literature values for high EQE devices nonetheless still show less than 5% power conversion efficiencies. The quantity limiting further efficiency improvement is the open circuit voltage ($V_{OC}$), whose importance as a key parameter has been largely overlooked, and thus paths to voltage improvement remain largely uninvestigated. Despite recent reports of high absorber material radiative efficiencies, an overwhelming majority of the reported $V_{OC}$ values for TMDC and other 2D semiconductors based photovoltaic devices are < 0.5 V[1, 5, 14-15] with record values of only ~0.8-0.9 V in split gated, in-plane homojunction devices[66, 68]. This implies a bandgap-$V_{oc}$ offset ($W_{oc} = E_g - V_{oc}$) > 0.8 V for most 2D semiconductor photovoltaic structures reported to date. A number of these reports have been for devices exhibiting a photovoltaic effect dominated by the Schottky barrier between the semiconductor and a metal or graphene contact.[37, 62, 88] Given that monolayer TMDC bandgaps lie generally in the range of 1.6-2.1 eV, whereas multilayer bandgaps range from 1.1-1.3 eV, there is still significant room to improve $V_{OC}$. For the case of monolayer absorbers, the large exciton binding energy due to the extreme 2D nature of carrier confinement, poses a challenge for the charge separation from bound excitons after absorption. To address the issue of charge separation and transport, it is useful to draw insights from concepts found in the literature for other excitonic devices, such as organic and dye sensitized solar cells.[89-93] To separate bound excitons, one either needs a junction in the active layer or carrier-selective contacts with built-in potential that exceeds the exciton binding energy. The large binding energy will nonetheless result in a voltage penalty.[94-95] Strategies that may enable the voltage penalty and exciton binding energy to be reduced include increasing the carrier concentration[96] or adding cladding layers with high static dielectric constants.[86, 97] Achieving high $V_{OC}$ therefore remains a critical hurdle towards achieving high efficiency photovoltaic devices from atomically thin semiconductor absorber layers. Key strategies to address the low $V_{OC}$ include achieving control over doping and band alignment for pn homojunctions and heterojunctions, in addition to optimizing the band alignments, bandgaps and conductivity of materials and interfaces used to form carrier selective contacts. Progress will likely require a systematic interdisciplinary effort combining concepts from chemistry, physics, and materials science to achieve this goal.

In the future, atomically thin materials will continue to garner attention for ultrathin and ultralight weight photovoltaics. However, no new photovoltaic technology likely to have a large impact unless it offers attributes than are either superior to those for existing Si photovoltaics or which can be usefully combined with Si photovoltaics, in order to widen the adoption of photovoltaics by improving efficiency and lowering cost. Therefore, achieving high power conversion efficiencies should remain a prime objective for atomically thin photovoltaics in 2D materials. Notable however, is that the bandgaps for monolayer TMDCs are almost ideally suited for high-efficiency photovoltaics in a two-junction tandem photovoltaic design featuring a Si bottom cell as seen in Figure 4 b, especially considering the near-unity PLQY values recently observed. In addition to tandem designs combining 2D semiconductors with Si photovoltaics, the inherently lightweight and flexible nature of atomically thin absorbers may enable use of these materials in mobile and portable power applications as well as building integrated photovoltaic applications where design of semitransparent photovoltaics is desirable. Also of interest are applications requiring high radiation hardness, such as space-based photovoltaics. The van der Waals bonded layered structure of TMDCs and 2D materials may facilitate easier integration with other substrates and active layers.

Aside from practical, high-efficiency photovoltaic applications, TMDCs and other 2D semiconductors such as layered hybrid organic-inorganic perovskites[98-99] are also of considerable fundamental scientific interest in light-matter interactions and energy conversion. The ability to create photovoltaic devices where photo-excited carrier transit distances are comparable to tunneling and hot carrier diffusion lengths presents opportunities to investigate and develop fundamentally novel mechanisms for energy conversion involving electromagnetic radiation. In particular, the presence of highly stable excitons at room temperature in free standing, oxide free, van der Waals layers represents a distinctive class of materials that possesses the radiative efficiency attributes of both direct gap inorganic semiconductors (GaAs/AlGaAs, GaN/InGaN quantum wells) and organic quantum confined semiconductors (small aromatic molecules, semiconducting polymers and carbon nanotubes). Further, the ability to electrostatically and dynamically tune the environment around these quantum-confined semiconductors, in order to influence their optical properties, presents new avenues for photonics and optoelectronics.

**Acknowledgements:** This work is part of the "Light-Material Interactions in Energy Conversion" Energy Frontier Research Center funded by the U.S. Department of Energy, Office of Science, Office of Basic Energy Sciences under award no. DE-SC0001293. D.J. and A.R.D., acknowledge additional support from the Space Solar Power project and the Resnick Sustainability Institute Graduate and Postdoctoral Fellowships. A.R.D. also acknowledges support in part from the Kavli Nanoscience Institute Postdoctoral Fellowship. J.W. acknowledges support from the National Science Foundation Graduate Research Fellowship under grant no. 1144469. All the authors acknowledge support of the Space Solar Power Initiative at Caltech funded by the Northrop Grumman Corporation.

**Figures:**

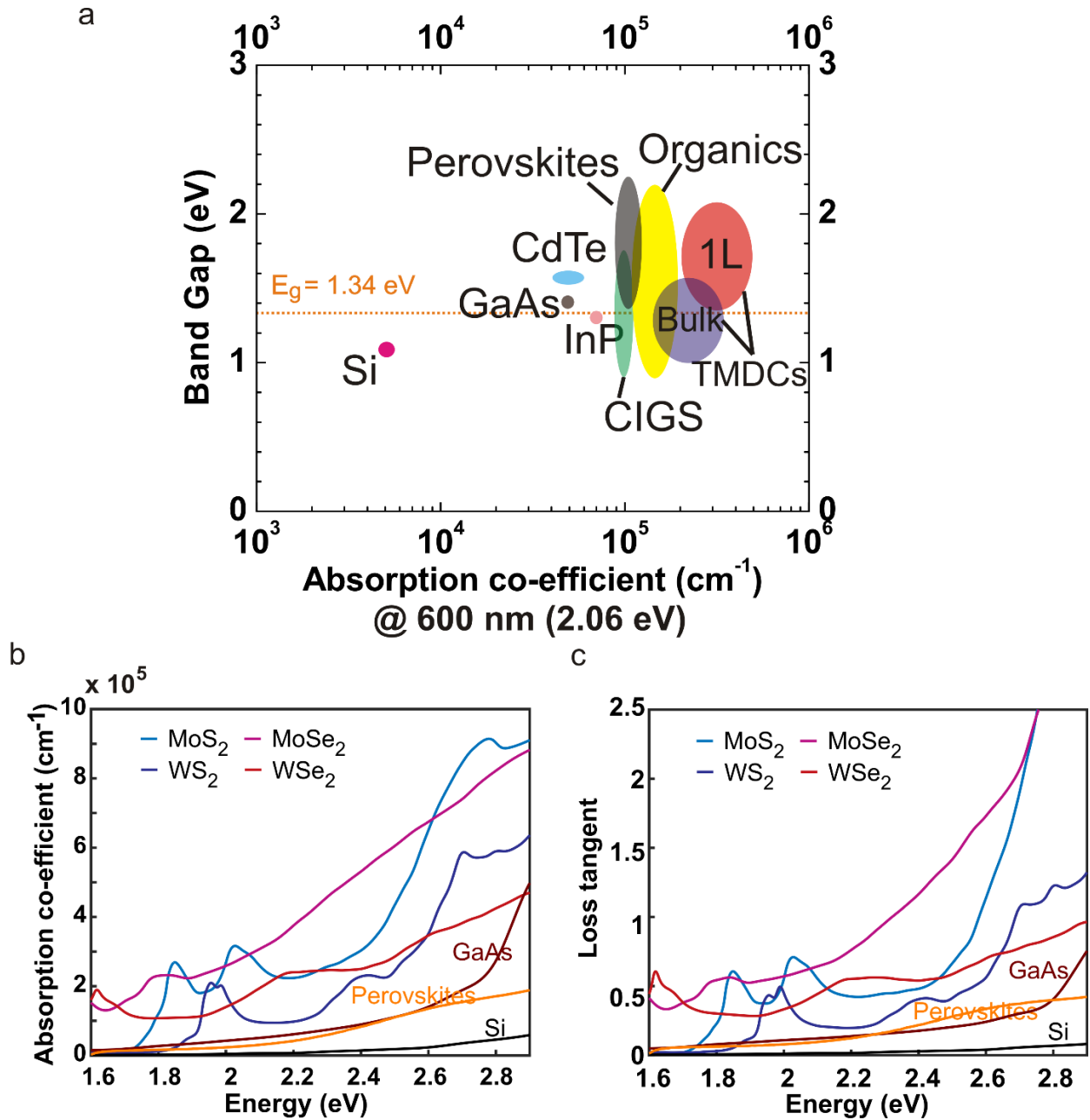

**Figure 1: Semiconductor absorption figures of merit for photovoltaic applications:** a. Comparison of energy bandgaps (eV) and absorption coefficients (cm$^{-1}$) for a variety of semiconductor materials used for commercial as well as research-scale photovoltaics. The TMDCs (both bulk and monolayers) of Mo and W have some of the highest absorption coefficients among known materials. b. Spectral absorption coefficient for selected photovoltaic materials, including Si and GaAs, as well the newly emerging methyl ammonium lead iodide perovskites (MAPbI$_3$) alongside the TMDCs. c. Loss tangent for the same materials in (b).

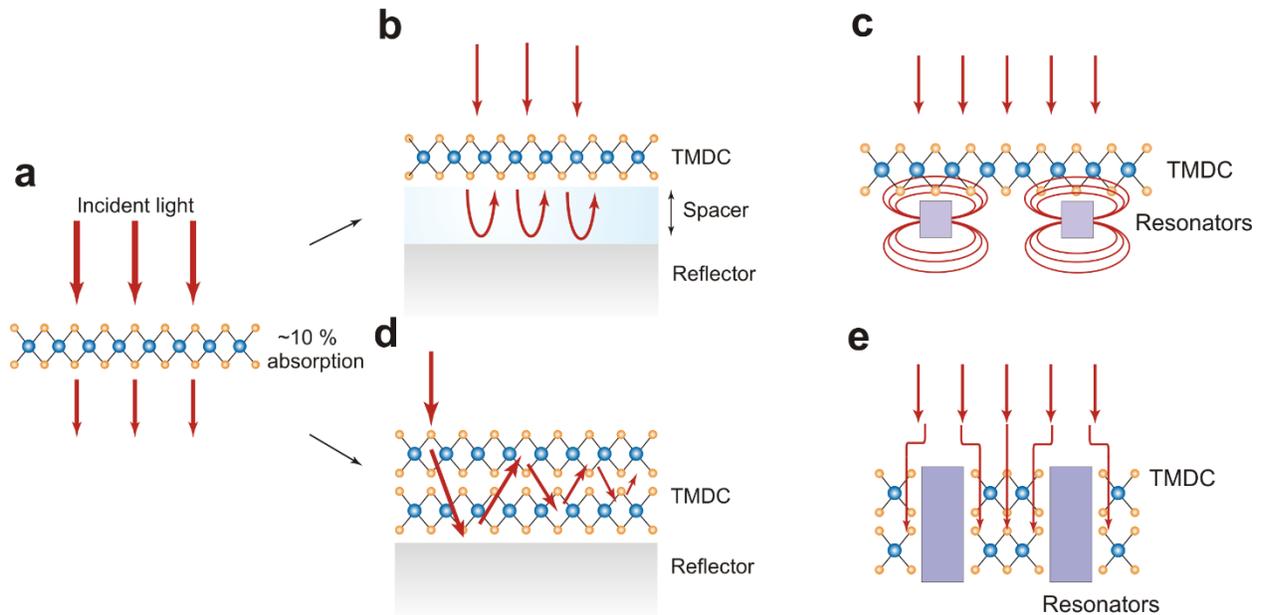

**Figure 2: Possible light trapping configurations for enhancing sunlight absorption:** a. A freestanding TMDC monolayer absorbs only a fraction of the incident sunlight (~10%), necessitating the use of light trapping techniques to increase the absorption. b. Monolayer absorber in a Salisbury screen-like configuration where the spacer thickness is ~$\lambda/4$ and the reflector is a low loss metal such as Ag, Au, or Al. c. Schematic of a TMDC monolayer coupled with resonators/antennas to enhance light absorption. d. Schematic of ultrathin, multilayer van der Waals absorber directly placed on a smooth reflective metal, where absorption is due to thin film interference. e. Resonantly absorbing nanometer scale antennas/resonators etched into a multilayer van der Waals material.

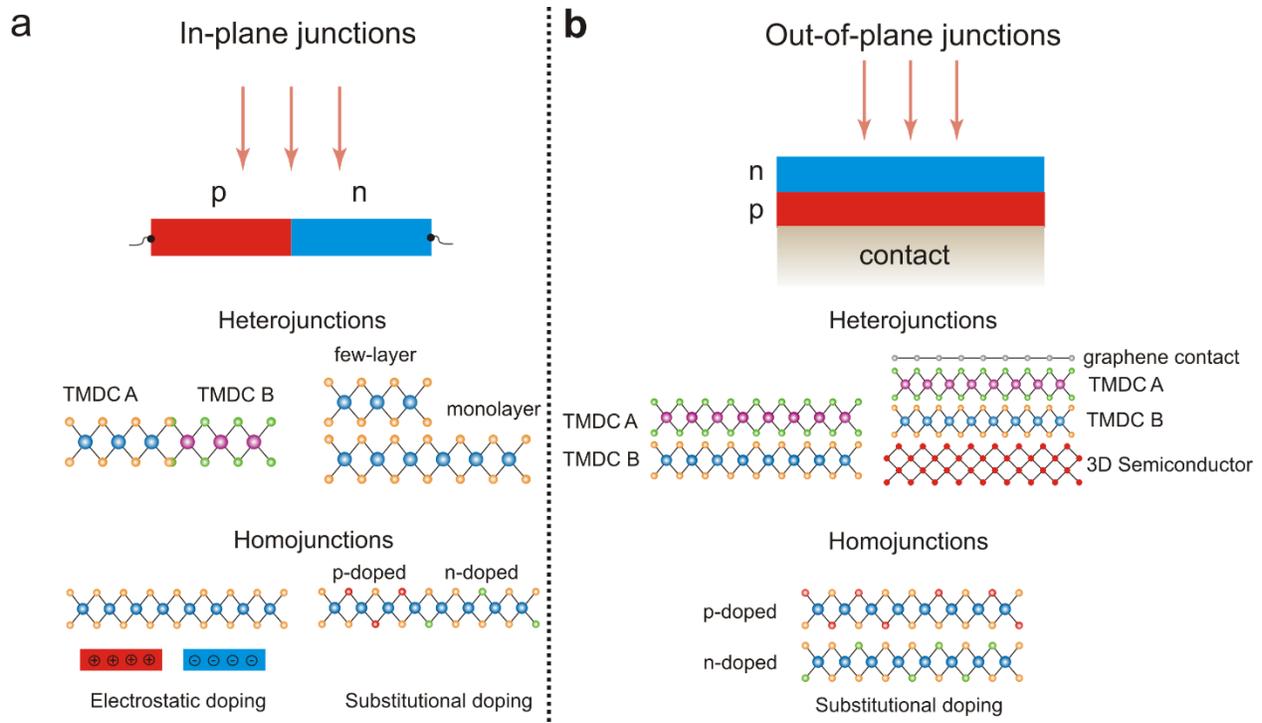

**Figure 3. Carrier collection schemes for Van der Waals materials and structures:** a. Schematic in-plane junction concepts for photovoltaic devices. Heterojunctions can be formed between two TMDC layers, as well as within the same TMDC material in which the thickness varies, since the bandgap is a thickness-dependent parameter in the ultrathin limit. Homojunctions can be created by electrostatic or modulation doping as well as substitutional doping. b. Schematic diagrams for out-of-plane junction concepts. The contacts between active layers are primarily van der Waals in nature. Heterojunctions can be formed by integrating two or more disparate TMDCs with different doping types and concentration. Van der Waals material heterostructures can also be integrated with conventional photovoltaic materials such as Si or III-V materials to make tandem cells. Graphene can effectively serve as transparent top contact material. Out-of-plane homojunctions can only be formed with substitutionally-doped layers stacked on top of one another.

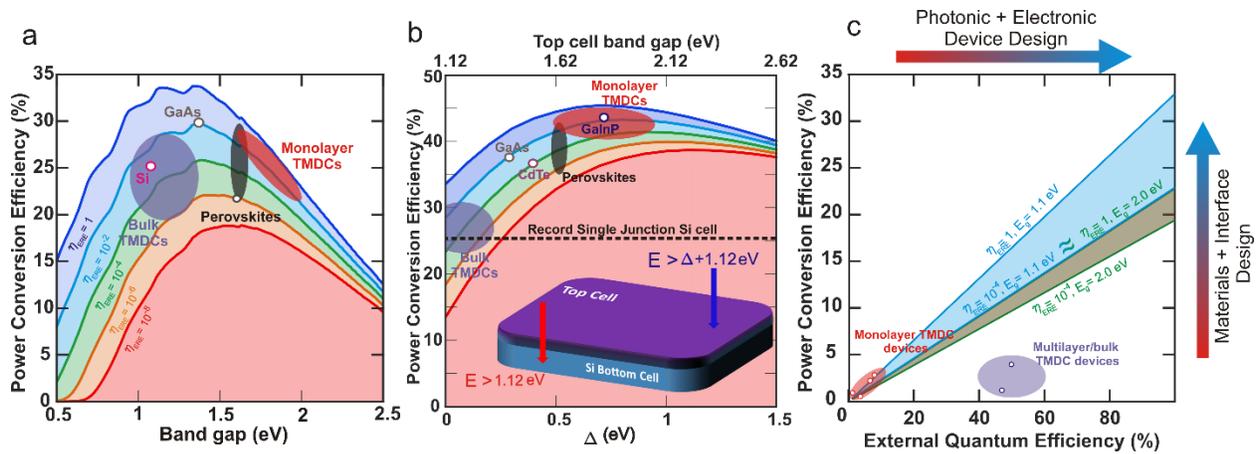

**Figure 4. Photovoltaic efficiency analysis: comparing TMDCs with established photovoltaic technologies**
a. photovoltaic efficiency of a single junction cell using a modified Shockley-Queisser detailed balance model that assumes a non-unity external radiative efficiency (ERE) of the semiconductor absorber layer. Some of the 2D material absorbers have been included, based on known or estimated values of ERE from PLQY reported or achieved in literature. b. Detailed balance power conversion efficiency estimates for a tandem cell structure with monocrystalline Si as the bottom cell. The plot colors correspond to varying ERE values, as depicted in a. The materials parameters are based on known ERE values[100] or record device performance.[101-103] c. Plot of efficiency as a function of external quantum efficiency (EQE) for two different values of ERE (1 and $10^{-4}$) for materials with bandgaps ~1.1 eV and 2 eV corresponding to upper and lower bounds in available TMDC bandgaps. The relevant 2D materials based devices have been appropriately mapped onto the plot based on literature reports.[48, 59, 61, 64, 66, 69]

## TOC Graphic

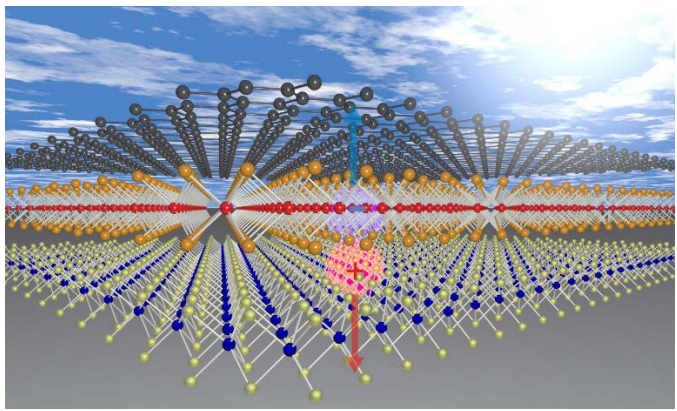